\documentclass[a4paper,12pt]{article}

\usepackage{graphics,amsmath,amssymb,epsfig}

\begin{document}

\title{
\bf Generalized models of unification of dark matter and dark energy}

\author{Neven \v{C}aplar\thanks{ncaplar@webmail.fizika.org}  and Hrvoje \v Stefan\v ci\'c$^{1,}$\thanks{shrvoje@thphys.irb.hr}}


\vspace{3 cm}
\date{
\centering
$^{1}$ Theoretical Physics Division, Rudjer Bo\v{s}kovi\'{c} Institute, \\
   P.O.Box 180, HR-10002 Zagreb, Croatia}


\maketitle

\abstract{A model of unification of dark matter and dark energy based on the modeling of the speed of sound as a function of the parameter of the equation of state 
is introduced. It is found that the model in which the speed of sound depends on the power of the parameter of the equation of state, $c_s^2=\alpha (-w)^{\gamma}$, 
 contains the generalized 
Chaplygin gas models as its subclass. An effective scalar field description of the model is obtained in a parametric form which in some cases can be translated 
into a closed form solution for the scalar field potential. A constraint on model parameters is obtained using the observational data on the Hubble parameter at different 
redshifts.}

\vspace{2cm}

\section{Introduction}

The expansion of the universe is one of the most fascinating phenomena that science has encountered so far. It has served as a rich source of information on the 
nature and the composition of the universe. One of recently established astonishing features of cosmic expansion is that it is currently undergoing a phase of 
acceleration \cite{R98,P99,S03,T04}. The source of this acceleration has not yet been unambiguously identified, although many proposals for 
its nature have been put forward (see \cite{Copeland,Saridakis} and references therein). The existence of 
dark energy, a mysterious component with negative pressure is still the most serious candidate. Another dark component of the universe, dark matter, seems to leave its 
imprint on astrophysical and cosmological scales, ranging from galaxies to galactic clusters and large scale structure in the universe.

The idea that both dark matter and dark energy are actually the manifestations of a single dark component is both natural and appealing. It appeared early in the 
literature and its the most acclaimed representative is probably the Chaplygin gas \cite{K01} as a model of unification of dark matter and dark energy \cite{Bi02}.  
The class of unifying DM-DE models is often referred to as quartessence \cite{Makler,Reis}.    
This phenomenologically introduced model can be motivated from string theory \cite{Bi02,O00,R00}. 
Its usually studied extension, the generalized Chaplygin gas model, was  first introduced in \cite{B02}. 
The agreement of the Chaplygin gas and its extensions with observations has been extensively 
tested, including the analyses with the supernova Ia data \cite{FGS02,C08}, CMB \cite{AFBC03}, observable Hubble parameter data \cite{Wu,F11} and large scale structure 
observations \cite{ST04,B03,F08,GK08,F10}, including the nonlinear evolution in structure formation \cite{Bilic2004} and 
gravastar formation \cite{Bilic2006Grav}. Different data can be combined to produce tighter parameter constrains such as in 
\cite{P10,BD03,CV09,LWY09}. Strong constraints on the generalized Chaplygin gas have been obtained that question its viability as a cosmological 
model distinguishable from the $\Lambda$CDM model. In order to better accommodate 
observational constrains, various unified models based on Chaplygin gas have been proposed, such as the modified Chaplygin gas model \cite{barrow,benaoum}, 
recently reviewed  and 
constrained in \cite{L08,D11,LXWL11} or hybrid Chaplygin gas leading to transient acceleration \cite{bilic2005trans}. The fact that perfect fluid model can be 
fully described by defining speed of sound equation has 
been used in \cite{X11}. The idea of DM-DE unification with non-canonical scalar fields has been recently studied in \cite{BB07,DFP12}. 
An interesting model called {\em Dusty Dark Energy}, recently introduced in \cite{Vikman}, achieves the DM-DE unification in the formalism of 
the $\lambda \varphi${\em -fluid}, resulting in the zero speed of sound and one scalar degree of freedom.
Other approaches to models of DM-DE unification that 
avoid the speed of sound problem are purely kinetic k-essence models \cite{S04,B07,Chimento} and tachyon models \cite{Padmanabhan,bilic2009tach}.

The structure of the paper is the following. After the introduction presented in this section, in the second section
 a general class of barotropic fluid models defined by the function $c_s^2$ is discussed. In the third section the model with the constant speed of sound is studied 
and in the fourth section the principal model of the paper, defined by $c_s^2(w)=\alpha (-w)^{\gamma}$ is introduced. The fifth section is focused on an effective 
representation of the model in terms of a minimally coupled scalar field. In the sixth section the comparison of the model prediction against the observational data on
the Hubble parameter at different redshifts is made and the constraints on the model parameters are presented. The seventh section closes the paper with the 
discussion and conclusions. The Appendix outlines an approach in which the solution with the piecewise constant speed of sound is used as an approximation of the 
dynamics of a fluid with a general dependence of $c_s^2$ on $w$.     

\section{The model}

The equation of state of a barotropic cosmic fluid can in general be written as an implicitly defined relation between the fluid pressure
 $p$ and its energy density $\rho$
\begin{equation}
\label{eq:EOS}
F(\rho,p)=0 \, .
\end{equation}
The parameter of the equation of state $w=p/\rho$ can be used to substitute the pressure, $p=w \rho$, so that the relation (\ref{eq:EOS}) 
becomes $G(\rho, w)=0$, with $G(\rho,w)=F(\rho,p)$. This relation implies that $\rho$ and $p$ can be considered as functions of $w$, i.e. $\rho=\rho(w)$ and 
$p=p(w)=w \rho(w)$. 
Strictly speaking, the inversion of the expression $G(\rho,w)=0$ may result in several solutions for $\rho(w)$ (and $p(w)$). In particular, for some value of 
$w$ there could be several values $\rho(w)$. The considerations presented below apply to each of these individual solutions.

The speed of sound waves in the barotropic cosmic fluid is defined as 
\begin{equation}
\label{eq:cs}
c_s^2=\frac{d p}{d \rho} \, .
\end{equation}
From (\ref{eq:EOS}) it follows
\begin{equation}
\label{eq:diffF}
\frac{\partial F}{\partial  \rho} d \rho + \frac{\partial F}{\partial p} d p = 0 \, ,
\end{equation}
which leads to 
\begin{equation}
\label{eq:cspart}
c_s^2=-\frac{\frac{\partial F}{\partial  \rho}}{\frac{\partial F}{\partial p}} \, .
\end{equation}
Inserting the relation $p=w \rho$ into (\ref{eq:diffF}) and using (\ref{eq:cspart}), one readily obtains 
\begin{equation}
\label{eq:rhow}
\frac{d \rho}{\rho} = \frac{d w}{c_s^2-w}\, .
\end{equation}
Combining this expression with the continuity equation for the fluid
\begin{equation}
\label{eq:cont}
d \rho+ 3 \rho (1+w) \frac{d a}{a}=0 \, ,
\end{equation}
one finally obtains
\begin{equation}
\label{eq:wdyn}
\frac{d w}{(c_s^2-w)(1+w)}=-3 \frac{d a}{a} = 3 \frac{d z}{1+z}\, .
\end{equation}
As $p$ and $\rho$ are functions of $w$, the expression for the speed of sound can be written as 
\begin{equation}
\label{eq:sound}
c_s^2=\frac{d p}{d \rho} = \frac{\frac{d p}{d w}}{\frac{d \rho}{d  w}} \, ,
\end{equation}
i.e. $c_s^2=c_s^2(w)$. Therefore, the expression (\ref{eq:wdyn}) is a dynamical law for the EOS parameter $w$.

The only unknown part in (\ref{eq:wdyn}) is the functional form of  $c_s^2(w)$. By its very definition there are some constraints on it, 
such as that it should be nonnegative and smaller than the speed of light squared, $c^2$. The modeling of cosmic fluid unifying dark matter and dark energy by
modeling $c_s^2$ as a function of $w$ may, therefore, be more suitable since the requirements on $c_s^2$ can be immediately built in.
Furthermore, the properties of the solutions of (\ref{eq:wdyn}) depend on the zeros of the function $c_s^2(w)-w$. In particular, the dynamics of $w(z)$ is confined to 
intervals determined by the zeros of $c_s^2(w)-w$ and $w=-1$. For some models with a variable speed of sound see \cite{bertacca2008,camera1,camera2,camera3}.   

In the following two sections we discuss specific choices for $c_s^2$ and discuss their relation with models previously studied in the literature.

\section{Constant speed of sound}

The simplest possibility is a constant speed of sound. This possibility was recently studied in \cite{X11}, see also \cite{affine,aviles,luongo,luongo2}. 
A direct inspection of (\ref{eq:wdyn}) shows that for the case 
of constant speed of sound $w$ is confined to one of the intervals: $(-\infty,-1)$, $(-1,c_s^2)$ and $(c_s^2,\infty)$.
For $c_s^2=\mathrm{const}$, the parameter of EOS evolves with the scale factor as
\begin{equation}
\label{eq:constw}
w=\frac{c_s^2 \frac{1+w_0}{c_s^2-w_0} (1+z)^{3(1+c_s^2)}-1}{ \frac{1+w_0}{c_s^2-w_0} (1+z)^{3(1+c_s^2)}+1} \, ,
\end{equation} 
with $w(0)=w_0$. From this relation it immediately follows
\begin{equation}
\label{eq:constrho}
\rho=\rho_0 \frac{c_s^2-w_0}{c_s^2-w}= \rho_0 \frac{c_s^2-w_0}{1+c_s^2} \left[ \frac{1+w_0}{c_s^2-w_0} (1+z)^{3(1+c_s^2)}+1 \right]
\end{equation}
and 
\begin{equation}
\label{eq:constp}
p= c_s^2 \rho - \rho_0 (c_s^2-w_0)= \rho_0 \frac{c_s^2-w_0}{1+c_s^2} \left[c_s^2 \frac{1+w_0}{c_s^2-w_0} (1+z)^{3(1+c_s^2)}-1 \right] \, .
\end{equation}
Finally we obtain
\begin{equation}
\label{eq:constsum}
\rho+p=\rho_0 (1+w_0) (1+z)^{3(1+c_s^2)} \, .
\end{equation}

For $-1 < w < c_s^2$, the energy density remains positive throughout the evolution of the universe. With the cosmic expansion, the pressure evolves from 
positive to negative values. In particular, the pressure changes sign at $1+z=(c_s^2 (1+w_0)/(c_s^2-w_0))^{-1/(3(1+c_s^2))}$.   
For $w_0<-1 $ the pressure is negative throughout the evolution of the universe, $p<0$. The energy density evolves from negative to positive values with the expansion 
and it changes sign at $1+z=(-(1+w_0)/(c_s^2-w_0))^{-1/(3(1+c_s^2))}$. For $w_0 > c_s^2$, the pressure is always positive, whereas the energy density evolves from 
positive to negative values with the expansion, with a change of sign at $1+z=((1+w_0)/(w_0-c_s^2))^{-1/(3(1+c_s^2))}$. 

The solution for $c_s^2=\mathrm{const}$ can be used as a building block for approximating general functional forms $c_s^2(w)$. The discussion of a piecewise 
constant approximation of the general $c_S^2$ redshift dependent function is given in the Appendix.

\section{Power law dependence: $c_s^2= \alpha (-w)^{\gamma}$}

 \label{power}

\begin{figure}[htp]
\epsfig{file=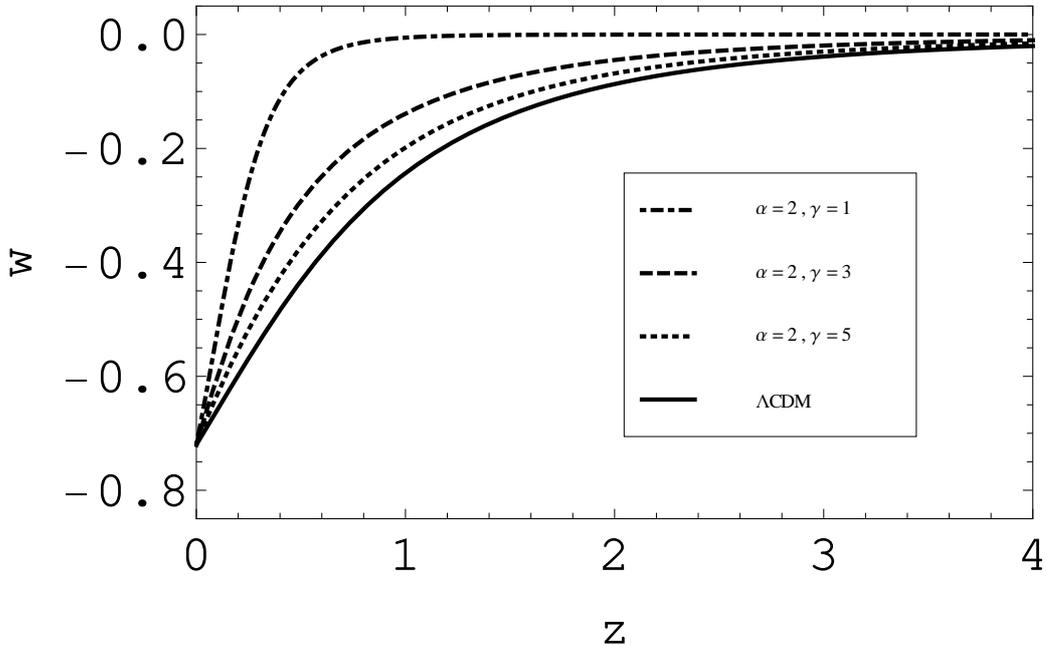,width=0.9\linewidth,clip=}
	\caption{The redshift dependence of the EOS parameter of the unified DM-DE component for several values of $(\alpha,\gamma)$ parameters. For all
parameter values the present value of the EOS parameter, $w_0$, is selected to match the value of the spatially flat $\mathrm{\Lambda CDM}$ model.} 
	\label{fig:w}
\end{figure}

\begin{figure}[htp]
\epsfig{file=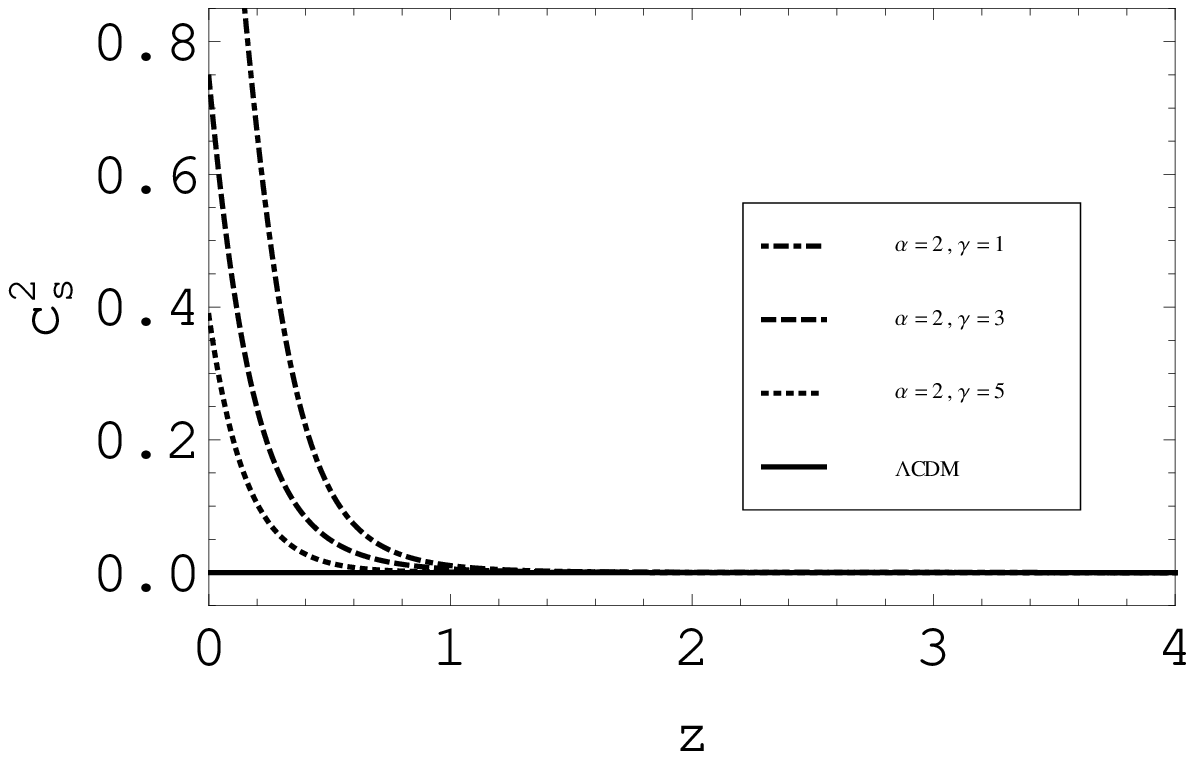,width=0.9\linewidth,clip=}
	\caption{The redshift dependence of the speed of sound squared, $c_s^2$, 
of the unified DM-DE component for several values of $(\alpha,\gamma)$ parameters. For all
parameter values the present value of the EOS parameter, $w_0$, is selected to match the value of the spatially flat $\mathrm{\Lambda CDM}$ model. } 
	\label{fig:cs}
\end{figure}

For a more general parametrization 
\begin{equation}
\label{eq:gen}
c_s^2= \alpha (-w)^{\gamma} \, , 
\end{equation}
the equation (\ref{eq:wdyn}) in general needs to be solved numerically to obtain $w=w(z)$. An example of such a numerical solution is presented in Fig \ref{fig:w}.
The redshift dependence of the speed of sound is presented in Fig. \ref{fig:cs}. We are primarily interested in the solutions in which $w$ is confined to the 
interval $(-1,0)$ and the model is properly defined for $w<0$.
For $\gamma > 1$ and $\alpha > 0$ the dynamics of $w$ is confined to one of the 
intervals $(-\infty,-1)$, $(-1,0)$, $(0,\infty)$. For $\gamma > 1$ and $\alpha < 0$, Eq. (\ref{eq:wdyn}) can be written as
\begin{equation}
 \label{eq:negalpha}
(1+z) \frac{d w}{d z} = -3 \alpha w (w+1) ((-w)^{\gamma-1}-(-w_{*})^{\gamma-1}) \, ,
\end{equation}
where $w_{*}=-(-1/\alpha)^{1/(\gamma-1)}$ is the additional zero of the denominator in (\ref{eq:wdyn}). For $\alpha < -1$, $w_{*}$ falls in the 
interval $(-1,0)$ which is of primary interest in this paper. Depending on the relation of $w_0$ and $w_{*}$, there are two distinct cases.
For $w_0 > w_{*}$, with the expansion $w(z)$ evolves from $w=0$ at early times towards $w_{*}$ which it asymptotically reaches in the infinity.
For $w_0 < w_{*}$ the expansion starts from $w=-1$ and asymptotically reaches $w_{*}$ in the infinity. In the latter case the model cannot serve as a
model of unification of dark matter and dark energy, but it could serve as a model of dark energy only.  

The EOS of the cosmic fluid with this parametrization of the speed of sound, however, can be calculated analytically. From the definition
\begin{equation}
\label{eq:genundef}
c_s^2=\frac{d p}{d \rho} = \alpha \left( -\frac{p}{\rho} \right)^{\gamma} \, , 
\end{equation}
for $\gamma \not = 1$, one readily obtains
\begin{equation}
\label{eq:rhogen}
\rho=\rho_0  \left[ \frac{\alpha+(-w_0)^{1-\gamma}}{\alpha+(-w)^{1-\gamma}} \right]^{1/(1-\gamma)} \, .
\end{equation}
For $\gamma=1$, the EOS becomes
\begin{equation}
\label{eq:gamma1}
p=A \rho^{-\alpha} \, .
\end{equation}

The result (\ref{eq:gamma1}) reveals that for $\gamma = 1$, the parametrization (\ref{eq:gen}) is equivalent to generalized Chaplygin gas, 
whereas for $\gamma=1, \alpha=1$ the EOS of Chaplygin gas is reproduced. Therefore,  (\ref{eq:gen}) represents a broad new class of unification models which 
contains the Chaplygin gas as its special case and the generalized Chaplygin gas as its subclass. The study of the model (\ref{eq:gen}) could potentially lead to 
deeper understanding of strong constraints appearing in the comparison of the Chaplygin gas and the generalized Chaplygin gas with the observational data.

For $\gamma=2$ and $\alpha \not = -1$, Eq. (\ref{eq:wdyn}) can be solved analytically and presented in a closed form:
\begin{equation}
 \label{eq:gamma2}
\frac{w_0}{w} \left( \frac{w+1}{w_0+1} \right)^{1/(\alpha+1)} \left( \frac{w-1/\alpha}{w_0-1/\alpha} \right)^{\alpha/(\alpha+1)} = (1+z)^3 \, .
\end{equation}

Furthermore, for a special case $\alpha=1$, explicit expressions for the evolution can be obtained. The function $w(z)$ then yields
\begin{equation}
 \label{eq:alpha1w}
w(z)=-\frac{1}{\sqrt{1+\frac{1-w_0^2}{w_0^2}(1+z)^6}} \, ,
\end{equation}
whereas the energy density and pressure read
\begin{equation}
 \label{eq:alpha1rho}
\rho(z)=-w_0 \rho_0 \frac{1+\sqrt{1+\frac{1-w_0^2}{w_0^2}(1+z)^6}}{-w_0+1}
\end{equation}
and
\begin{equation}
 \label{eq:alpha1p}
p(z)=\rho_0 w_0 \frac{1+\frac{1}{\sqrt{1+\frac{1-w_0^2}{w_0^2}(1+z)^6}}}{-w_0+1} \, ,
\end{equation}
respectively.

Finally, for $\alpha=-1/2$ Eq. (\ref{eq:gamma2}) can also be presented as an explicit expression for $w(z)$:
\begin{equation}
 \label{eq:rhominus0.5}
w(z)=-1+\frac{1}{\sqrt{1+\frac{1-(w_0+1)^2}{(w_0+1)^2}(1+z)^{-3}}} \, .
\end{equation}
The energy density and pressure are given by the expressions
 \begin{equation}
 \label{eq:wminus0.5}
\rho(z)=\rho_0 \frac{-w_0}{w_0+2} \frac{\sqrt{1+\frac{1-(w_0+1)^2}{(w_0+1)^2}(1+z)^{-3}}+1}{\sqrt{1+\frac{1-(w_0+1)^2}{(w_0+1)^2}(1+z)^{-3}}-1}
\end{equation} 
and 
\begin{equation}
 \label{eq:pminus0.5}
p(z)=\rho_0 \frac{w_0}{w_0+2} \left(1+\frac{1}{\sqrt{1+\frac{1-(w_0+1)^2}{(w_0+1)^2}(1+z)^{-3}}} \right) \, .
\end{equation}

Once the expressions for $\rho(z)$, $p(z)$ and $w(z)$ are available, the expression for $H(z)$ can be readily obtained. A straightforward integration would then yield
$a(t)$, the scale factor as a function of the cosmic time, which represents the full dynamical information. As the expressions for 
dynamical quantities in terms of redshift are sufficient for the description of the transition between the DM regime and DE regime and for the comparison 
with the observational data, we do not present numerical solutions for $a(t)$.

\section{Scalar field representation}

\label{sec:scalar}

If the cosmic fluid with the speed of sound (\ref{eq:gen}) is the dominant component in the universe, in a spatially flat FLRW universe the Hubble parameter is 
\begin{equation}
\label{eq:H2}
H^2=\frac{8 \pi G}{3} \rho \, .
\end{equation}
In this section we focus on finding the effective description of the cosmic dynamics with the cosmic fluid defined by (\ref{eq:gen}) in terms of a 
minimally coupled scalar field $\varphi$ with a potential $V(\varphi)$. 
For an effective representation of the cosmic fluid in terms of the scalar field, the expressions for $\rho$ and $p$ are
\begin{equation}
\label{eq:rhophi}
\rho=\frac{1}{2} \dot{\varphi}^2+V(\varphi) 
\end{equation}
and
\begin{equation}
\label{eq:pphi}
p=\frac{1}{2} \dot{\varphi}^2-V(\varphi) \, . 
\end{equation}

The time derivative of the scalar field can be expressed as $\dot{\varphi}^2=\rho+p=(1+w) \rho(w)$ from (\ref{eq:rhophi}) and (\ref{eq:pphi}). On the other hand
\begin{equation}
\label{eq:dotphi}
\dot{\varphi}=\frac{ d \varphi}{d w} \frac{ d w}{d t} = \frac{ d \varphi}{ d w} (-3) H (c_s^2-w)(1+w) \, .
\end{equation}
Combining these two expressions for $\dot{\varphi}$ and using (\ref{eq:H2}), the following equation is obtained
\begin{equation}
\label{eq:phieq}
\sqrt{\frac{8 \pi G}{3}} d \varphi = \mp \frac{1}{3} \frac{1}{\sqrt{1+w}(c_s^2-w)} d w \, .
\end{equation}
Integration of this equation leads to $\varphi=\varphi(w)$. In a similar way, combining (\ref{eq:rhophi}) and (\ref{eq:pphi}) the scalar field potential is
\begin{equation}
\label{eq:Vphi}
V(\varphi)=\frac{1}{2} (1-w) \rho(w) \, .
\end{equation} 
Therefore, assuming that (\ref{eq:phieq}) is analytically integrable, the scalar field representation is available in the parametric form as 
$\varphi=\varphi(w)$ and $V(\varphi)=V(w)$. Even in the case when it is not possible to analytically integrate (\ref{eq:phieq}), Eqs (\ref{eq:phieq}) and 
(\ref{eq:Vphi}) provide a direct procedure (including numerical integration for each value of $w$) for the parametrically defined pair of quantities
$\varphi(w),V(w)$. In the remainder of this section we consider an analytical scalar field reconstruction for several specific 
examples and use the abbreviation $\phi=\sqrt{\frac{8 \pi G}{3}} \varphi$.

\subsection{$c_s^2=\alpha$}

For the case of constant sound speed with $\alpha>-1$, the equations (\ref{eq:constrho}) and (\ref{eq:Vphi}) yield
\begin{equation}
\label{eq:constVw}
V(\phi)=\frac{1}{2} \rho_0 (\alpha-w_0) \frac{1-w}{\alpha-w} \, .
\end{equation}
On the other hand, Eq. (\ref{eq:phieq}) is readily integrated to 
\begin{equation}
\label{eq:constphi}
\phi-\phi_0= \mp \frac{1}{3 \sqrt{1+\alpha}} \left( \ln \frac{\sqrt{1+w} + \sqrt{1+\alpha}}{\sqrt{1+w_0} + \sqrt{1+\alpha}} - 
\ln \frac{\sqrt{1+w} - \sqrt{1+\alpha}}{\sqrt{1+w_0} - \sqrt{1+\alpha}}\right) \, .
\end{equation}
The parameter of EOS is then 
\begin{equation}
 \label{eq:wody}
w=(1+\alpha) \left( \frac{y+1}{y-1} \right)^2 -1 \, ,
\end{equation}
where
\begin{equation}
 \label{eq:defy}
y=\frac{\sqrt{1+w_0} + \sqrt{1+\alpha}}{\sqrt{1+w_0} - \sqrt{1+\alpha}} e^{\mp 3 \sqrt{1+\alpha} (\phi-\phi_0)} \equiv - e^{\mp 3 \sqrt{1+\alpha} (\phi-\tilde{\phi})} \, .
\end{equation}

Combining (\ref{eq:constVw}), (\ref{eq:wody}) and (\ref{eq:defy}) the scalar field potential is
\begin{equation}
 \label{eq:constVphi}
V(\phi) = \frac{1}{2} \rho_0 (\alpha-w_0) \left[1+ \frac{1-\alpha}{1+\alpha} \left( \cosh \left( \frac{3}{2} \sqrt{1+\alpha} (\phi-\tilde{\phi}) \right) \right)^2 \right] \, .
\end{equation}

\subsection{$c_s^2=-\alpha w$}

In this section we consider the sound speed linearly dependent on the EOS parameter. As already shown in section \ref{power}, such a dependence is characteristic of
the generalized Chaplygin gas. In this subsection we reconstruct the scalar potential for the generalized Chaplygin gas using the method of parametric definition 
of $\phi$ and $V(\phi)$ in terms of $w$. The integration of (\ref{eq:phieq}) gives
\begin{equation}
 \label{eq:lin1}
\pm 3 (1+\alpha) (\phi-\phi_0) = \ln \frac{\sqrt{1+w}-1}{\sqrt{1+w_0}-1} - \ln \frac{\sqrt{1+w}+1}{\sqrt{1+w_0}+1} \, ,
\end{equation}
which results in 
\begin{equation}
 \label{eq:lin2}
w=\left( \frac{1+y}{1-y} \right)^2-1 \, ,
\end{equation}
where
\begin{equation}
 \label{eq:lin3}
y=\frac{\sqrt{1+w_0}-1}{\sqrt{1+w_0}+1} e^{\pm 3 (1+\alpha) (\phi-\phi_0)} \equiv - e^{\pm 3 (1+\alpha) (\phi-\tilde{\phi})} \, .
\end{equation}
The scalar field potential is 
\begin{equation}
 \label{eq:lin4}
V(\phi)=\frac{1}{2} \rho_0 (1-w) \left( \frac{w}{w_0} \right)^{-1/(1+\alpha)} \, . 
\end{equation}
The combination of (\ref{eq:lin2}), (\ref{eq:lin3}) and (\ref{eq:lin4}) results in the potential
\begin{eqnarray}
 \label{eq:lin5}
V(\phi)&=&\frac{1}{2} \rho_0 \left( -\frac{1}{w_0} \right)^{-1/(1+\alpha)} \left[ \left( \cosh \left( \frac{3}{2} (1+\alpha) (\phi-\tilde{\phi}) \right) 
\right)^{2/(1+\alpha)} \right. \nonumber \\
&+&  \left. \left( \cosh \left( \frac{3}{2} (1+\alpha) (\phi-\tilde{\phi}) \right) 
\right)^{-2 \alpha/(1+\alpha)} \right] 
\end{eqnarray}
for the generalized Chaplygin model. Specifying $\alpha=1$ we obtain the potential for the Chaplygin gas
\begin{equation}
 \label{eq:lin6}
V(\phi)=\frac{1}{2} \rho_0 (-w_0)^{1/2} \left( \cosh (3(\phi-\tilde{\phi})) + \frac{1}{\cosh (3(\phi-\tilde{\phi}))} \right) \, .
\end{equation}

\subsection{$c_s^2=\alpha (-w)^2$}

\begin{figure}[htp]
\epsfig{file=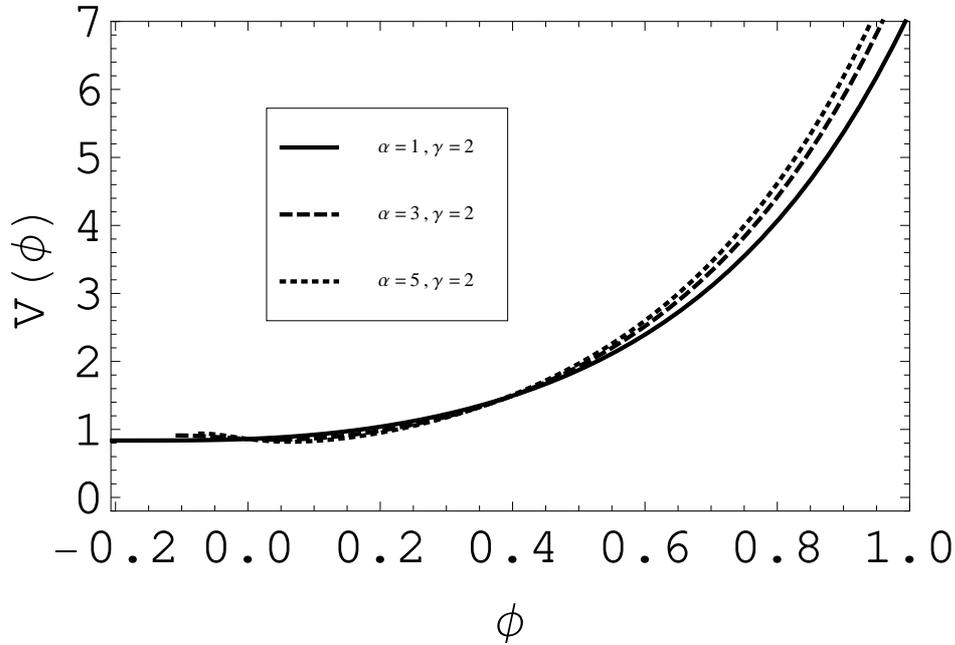,width=0.9\linewidth,clip=}
	\caption{The effective scalar field potential for several values of parameters $(\alpha,\gamma)$. The potential $V(\phi)$ is presented in units of $\rho_0$, using
$\phi_0=0$ and the $+$ sign in (\ref{eq:quad1}). } 
	\label{fig:potential}
\end{figure}

For $\gamma=2$ the integration of (\ref{eq:phieq}) leads to 
\begin{equation}
 \label{eq:quad1}
\phi =\phi_0 \mp \frac{1}{3} \ln \left[ \frac{\sqrt{1+w} +1}{\sqrt{w_0+1}+1}  \frac{\sqrt{1+w_0} -1}{\sqrt{w+1}-1} 
\left( \frac{\sqrt{1+w_0} +\beta}{\sqrt{w+1}+\beta}\right)^{1/\beta}  \left( \frac{\sqrt{1+w} -\beta}{\sqrt{w_0+1}-\beta}\right)^{1/\beta} \right] \, , 
\end{equation}
where $\beta = \sqrt{(\alpha+1)/\alpha}$. The scalar field potential has the form
\begin{equation}
 \label{eq:quad2}
V(\phi)=\frac{1}{2} \rho_0 \frac{w_0}{\alpha w_0-1} \frac{(1-w)(\alpha w-1)}{w} \, .
\end{equation}
Equations (\ref{eq:quad1}) and (\ref{eq:quad2}) provide the solution for the scalar field potential in a parametric form. For $\gamma > 1$ the evolution of 
the EOS parameter $w$ is confined to the interval $(-1,0)$. Taking the values for $w$ from this interval the plots of $V(\phi$) are readily obtained. The functional
forms of $V(\phi)$ for several values of $\alpha$ are depicted in Fig. \ref{fig:potential}.

\section{Numerical analysis and the comparison with the observational data}

In this section we perform a numerical analysis in order to find constraints on the model parameters from the 
comparison with the observational data. We use the measurements of the Hubble parameter at different redshifts obtained from the 
passively evolving galaxies \cite{JL02} and baryon acoustic oscillations (BAO). The principal idea of the former measurement technique is that for a pair of passively evolving galaxies at close 
redshifts, the differential of their age can be determined. Then from the expression $H(z)=-\frac{1}{1+z} \frac{d z}{d t}$ the value of $H(z)$ can be calculated.
This method has been recently successfully used to constrain the parameters of the generalized Chaplygin gas \cite{F11}. To constrain the 
model parameters $\alpha$ and $\gamma$, we use the $H(z)$ data reported in \cite{ZML10}.  

\begin{figure}[htp]
\epsfig{file=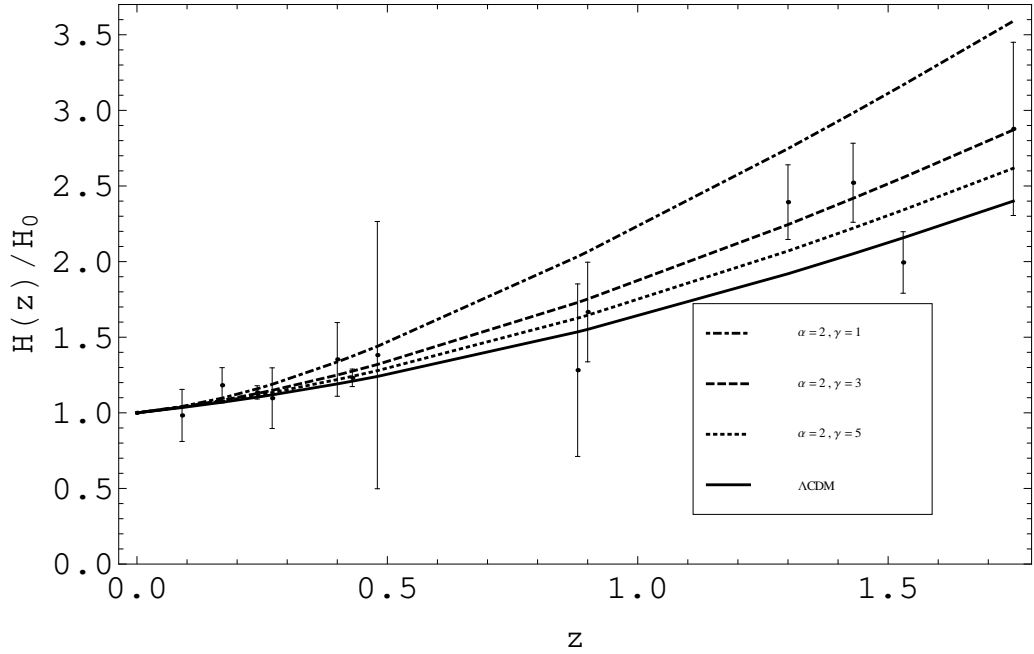,width=0.9\linewidth,clip=}
	\caption{The redshift dependence of the Hubble parameter for several values of $(\alpha,\gamma)$ parameters with $\Omega_b^0=0.0458$ and $w_0=-0.76$. The 
observed values for the Hubble parameters are taken from $\cite{ZML10}$ and the value for the present value of the Hubble parameter is 
$H_0=70.2 \pm 1.4 \mbox{ km }\mbox{s}^{-1} \mbox{ Mpc}^{-1}$ \cite{wmap7}.} 
	\label{fig:H}
\end{figure}

We restrict our analysis to the spatially flat Robertson-Walker metric motivated by the inflationary expansion and data on CMB anisotropies 
\cite{wmap7}. Furthermore, we fix
the share of baryons in the present energy density to $\Omega_b^0=0.0458$ \cite{wmap7}, in accordance with primordial nucleosynthesis requirements. The expression for the Hubble 
parameter then becomes
\begin{equation}
 \label{eq:Hflat}
H^2=\frac{8 \pi G}{3} (\rho_b+\rho) \, ,
\end{equation}
where $\rho_b$ denotes the energy density of baryons evolving with redshift as $\rho_b(z)=\rho_{b,0} (1+z)^3$ and $\rho$ denotes the energy density of the unified 
DM-DE component evolving with redshift as $\rho(z)=\rho_0 f(z)$, where (\ref{eq:rhogen}) gives
\begin{equation}
 \label{eq:fodz}
f(z)=\left[ \frac{\alpha+(-w_0)^{1-\gamma}}{\alpha+(-w(z))^{1-\gamma}} \right]^{1/(1-\gamma)} \, .
\end{equation}
The expression for the Hubble parameter then becomes
\begin{equation}
 \label{eq:Hodz}
H(z)=H_0 \left( \Omega_b^0 (1+z)^3 + (1-\Omega_b^0) f(z) \right)^{1/2} \, ,
\end{equation}
where $H_0= 100 h \mbox{ km }\mbox{s}^{-1} \mbox{ Mpc}^{-1}$ is the present value of the Hubble parameter. The functions $H(z)/H_0$ for different values of $(\alpha,\gamma)$ 
along with the observational data are presented in Fig. \ref{fig:H}.
The $\chi^2$ function is calculated according to the expression  
\begin{equation}
\label{eq:chi2}
\chi^{2}= \sum_{i}  \frac{(H^{th}(z_{i})-H^{obs}(z_{i}))^2}{\sigma^{2}_{i}}  \, ,
\end{equation}
with the observational data for $H(z)$ taken from \cite{ZML10}. Finally, from (\ref{eq:chi2}) the probability is calculated as
\begin{equation}
P = A e^{-\chi^{2}/2} \, ,
\end{equation}
where the symbol $A$ denotes the normalization. 

\begin{figure}[htp]
\epsfig{file=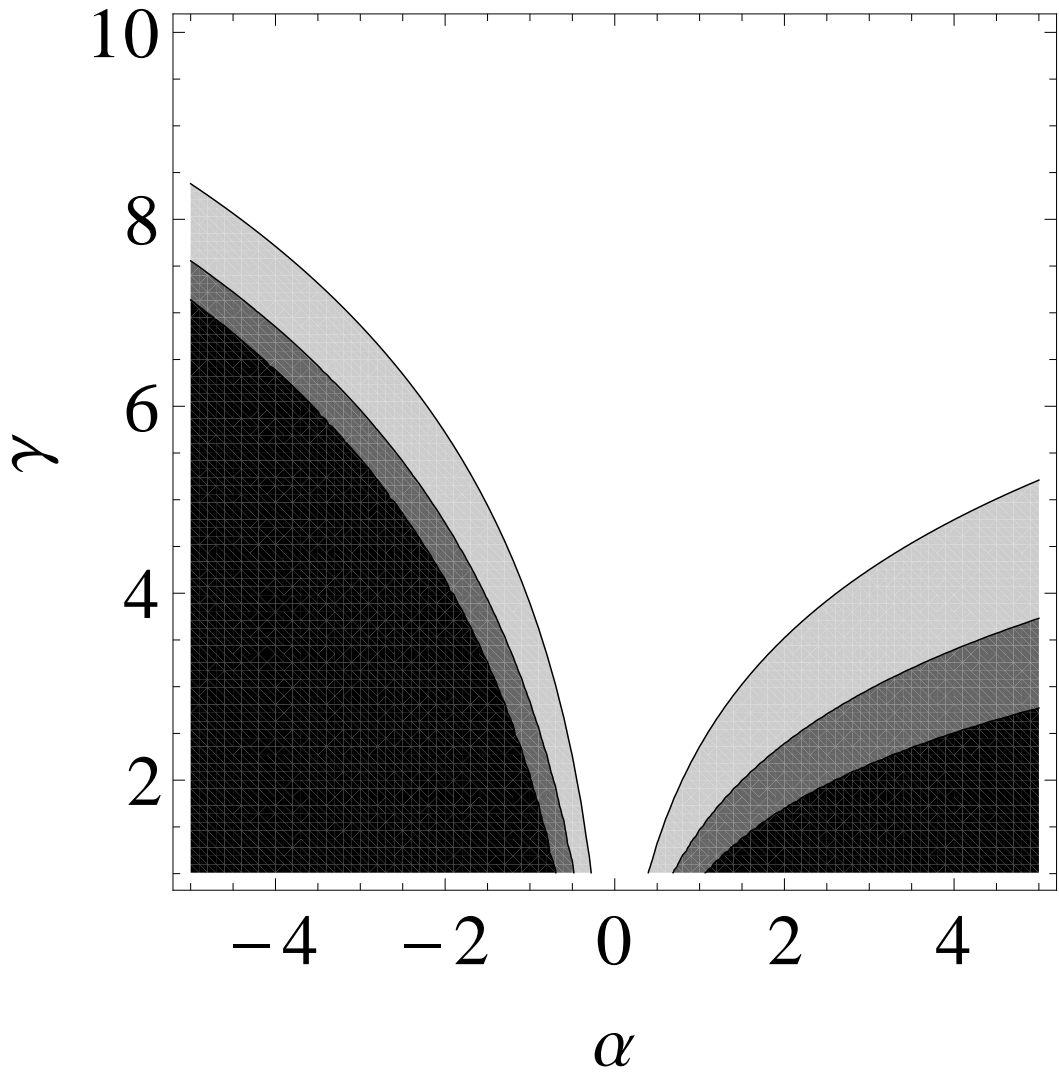,width=0.9\linewidth,clip=}
	\caption{The $68.3\%$ (white), $95.4\%$ (light gray) and $99.7\%$ (gray) probability contours in the $(\alpha,\gamma)$ plane after marginalizing over $h$.} 
	\label{fig:chi2}
\end{figure}

Although the theoretically preferred region for $\alpha$ is $(0,1)$, we run our model for a broader range of parameters $-5<\alpha<5$ and $1< \gamma < 10$. While we marginalize over the parameter $h$, we select 
$w_0=-\Omega_{\Lambda}^{0}/(1-\Omega_b^0)=-0.76$ (where $\Omega_{\Lambda}^{0}$ and $\Omega_b^0$ refer to percentages of the cosmological constant
and the baryonic matter in the $\mathrm{\Lambda}$CDM model \cite{wmap7}) to obtain constraints 
on the parameters $(\alpha,\gamma)$ which produce the present total EOS consistent with the value of the $\mathrm{\Lambda}$CDM model. 
The $(68.3\%, 95.4\%, 99.7\%)$ 
contours of the marginalized probability density are given in Fig \ref{fig:chi2}. They are calculated as curves for which $\Delta \chi^2=(2.30, 6.17, 11.8)$
where $\Delta \chi^2$ denotes the difference of $\chi^2$ at some point and its minimal value \cite{verde,havens}. 
From the figure it is clear that the allowed interval of $\alpha$ grows with the increase of $\gamma$.

\begin{figure}[htp]
\epsfig{file=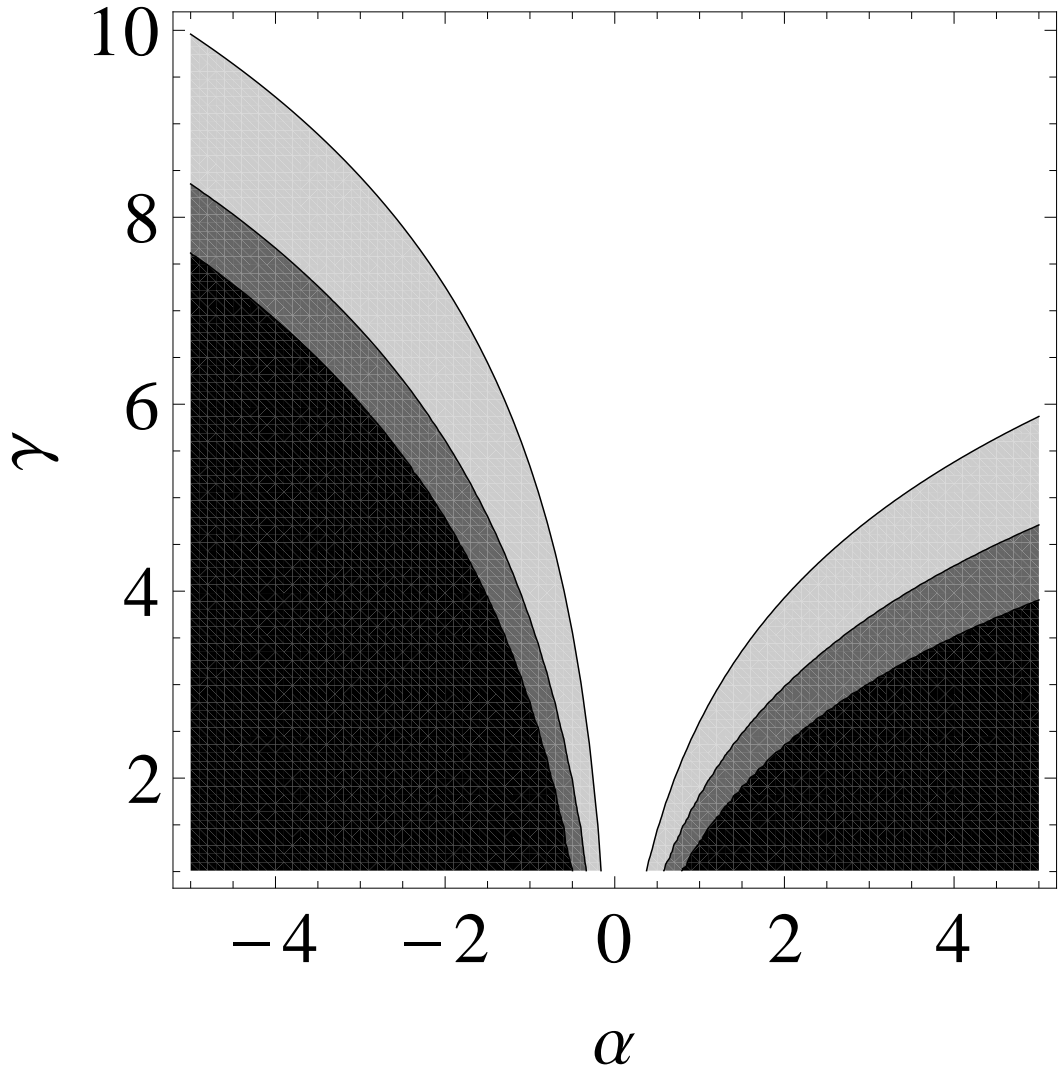,width=0.9\linewidth,clip=}
	\caption{The $68.3\%$ (white), $95.4\%$ (light gray) and $99.7\%$ (gray) probability contours in the $(\alpha,\gamma)$ plane for $h=0.702$ \cite{wmap7}.} 
	\label{fig:chi2H0}
\end{figure}

The contour plot for the $(68.3\%, 95.4\%, 99.7\%)$ probability intervals when no marginalization over $h$ is performed and the value for the Hubble parameter is taken 
from \cite{wmap7} to be $h=0.702$ is presented in Fig. \ref{fig:chi2H0}. From figures \ref{fig:chi2} and \ref{fig:chi2H0} it is evident that the probability contours for the 
$(\alpha,\gamma)$ parameters with and without marginalization are quite similar.  

\section{Discussion and conclusions}

The analysis presented in the preceding section serves primarily as an orientation regarding the allowed part of the parametric $(\alpha,\gamma)$ space. A more complete
analysis should consider the growth of inhomogeneities and make the comparison of the model predictions with other observational data, 
such as the data on the matter power spectrum, supernovae of the type Ia and cosmic microwave background. Although preliminary analyses in this direction have already been made 
\cite{CaplarDipl}, a detailed analysis of observational data is left for future work \cite{Prep}. 

From Figs \ref{fig:chi2} and \ref{fig:chi2H0} it is evident that for larger values of the exponent $\gamma$, a wider interval of the coefficient $\alpha$ is allowed. A similar conclusion 
follows from the preliminary analysis of the matter power spectrum \cite{CaplarDipl}. This feature can be understood on qualitative grounds using the following 
argumentation. For large values of $\gamma$, as long as the EOS parameter is not far from $w=0$, the speed of sound of the dark component remains small and 
suppressed by the large exponent $\gamma$. This fact prevents the formation of the sound horizon and its adverse effects on structure formation.

The generalized model of unification of dark matter and dark energy introduced in this paper opens a novel perspective on Chaplygin gas and its 
modifications and generalizations. The model defined by the relation $c_s^2=\alpha (-w)^{\gamma}$ encompasses both the Chaplygin gas (for $\gamma=1$, $\alpha=1$) and 
the generalized Chaplygin gas (for $\gamma=1$) as a specific subclass. It could therefore serve as a wider framework for the analysis how much these models need to 
be extended to satisfy the constraint from the observational data. From the modeling perspective, the crucial element of our model is the relation between the 
speed of sound, $c_S^2$, and the parameter of the EOS, $w$. This relation connects the quantity governing the growth of inhomogeneities with the quantity determining
the global evolution of the universe. This feature might allow easier and more direct transformation of the phenomenological knowledge acquired from the data into workable 
models of the dark sector. In particular, an adaptive model assuming piecewise constant values of the speed of sound in consecutive redshift intervals is presented in 
the Appendix.

A particular challenge for the future research is finding a microscopic explanation of the dependence $c_S^2(w)$. Here the corresponding microscopic models for 
the Chaplygin gas might serve as a good starting point.  
In particular, the method of representing the evolution in terms of minimally coupled scalar fields, presented in
section \ref{sec:scalar} could provide useful information on such microscopic models.

\vspace{0.5cm}

{\bf Acknowledgements.} H. \v{S}. acknowledges useful discussions with D. Huterer in the early phase of this work. 
The authors would like to thank N. Bili\'{c} for useful comments on the manuscript. This work was
supported by the Ministry of Education, Science and Sports of the Republic of Croatia 
under the contract No. 098-0982930-2864.

\section*{Appendix}

\label{app1}

In the comparison of models where the sound speed of the cosmic fluid is tested against the observational data, the goal is to test as general functional 
form $c_S^2(w)$ as possible. As this is in general quite a formidable task, a powerful approximation to the general case is to 
assume that $c_S^2$ is piecewise constant,
i.e. that for every interval of redshift $[z_{i-1},z_i)$, with $z_0=0$, the speed of sound is constant, i.e. $c_S^2=c_{S,i}^2$. In this approach the parameters are 
borders of the intervals $z_i$ and the constant sound speed values $c_{S,i}^2$.  
An example of this simple, but powerful parametrization was given in \cite{cooray} for the case of $w(z)$ function. 
The equations (\ref{eq:constw}), (\ref{eq:constrho}) and (\ref{eq:constp}) in
the redshift interval $[z_{i-1},z_i)$ become
\begin{equation}
 \label{eq:wi}
w(z)=\frac{c_{s,i}^2 \frac{1+w_{i-1}}{c_{s,i}^2-w_{i-1}} \left( \frac{1+z}{1+z_{i-1}} \right)^{3(1+c_{s,i}^2)}  -1 }{1+
\frac{1+w_{i-1}}{c_{s,i}^2-w_{i-1}} \left( \frac{1+z}{1+z_{i-1}} \right)^{3(1+c_{s,i}^2)}} \, ,
\end{equation}
\begin{equation}
\label{eq:rhoi} 
\rho(z)=\rho_{i-1} \frac{c_{s,i}^2-w_{i-1}}{1+c_{s,i}^2} \left[ 1+
\frac{1+w_{i-1}}{c_{s,i}^2-w_{i-1}} \left( \frac{1+z}{1+z_{i-1}} \right)^{3(1+c_{s,i}^2)} \right] 
\end{equation}
and
\begin{equation}
\label{eq:pi} 
p(z)=\rho_{i-1} \frac{c_{s,i}^2-w_{i-1}}{1+c_{s,i}^2} 
\left[ c_{s,i}^2 \frac{1+w_{i-1}}{c_{s,i}^2-w_{i-1}} \left( \frac{1+z}{1+z_{i-1}} \right)^{3(1+c_{s,i}^2)}  -1 \right] \, .
\end{equation}
The relations between the parameters $w_i$ and $\rho_i$ in neighboring intervals are given by relations
\begin{equation}
\label{eq:connectw} 
w_i \equiv w(z_i)=\frac{c_{s,i}^2 \frac{1+w_{i-1}}{c_{s,i}^2-w_{i-1}} \left( \frac{1+z_i}{1+z_{i-1}} \right)^{3(1+c_{s,i}^2)}  -1 }{1+
\frac{1+w_{i-1}}{c_{s,i}^2-w_{i-1}} \left( \frac{1+z_i}{1+z_{i-1}} \right)^{3(1+c_{s,i}^2)}}
\end{equation}
and 
\begin{equation}
\label{eq:connectrho} 
\rho_i \equiv \rho(z_i)=\rho_{i-1} \frac{c_{s,i}^2-w_{i-1}}{1+c_{s,i}^2} \left[ 1+
\frac{1+w_{i-1}}{c_{s,i}^2-w_{i-1}} \left( \frac{1+z_i}{1+z_{i-1}} \right)^{3(1+c_{s,i}^2)} \right] \, .
\end{equation}

The expression (\ref{eq:rhoi}) can be used, e.g. for the comparison of the model with piecewise constant speed of sound against the SN Ia data.

\end{document}